\documentclass[10pt,twocolumn]{revtex4-1}
\usepackage{bm}
\usepackage{hyperref}
\usepackage{makeidx}
\usepackage{amsmath}
\usepackage{graphicx}
\usepackage{dcolumn}
\usepackage{color}
\usepackage{graphics}
\usepackage{amsfonts}
\usepackage{amssymb}
\usepackage{epsfig}
\usepackage{float}
\usepackage[british,UKenglish,USenglish,english,american]{babel}
\usepackage{epstopdf}
\usepackage{graphics, subfigure}
\usepackage{array}

\begin{document}

\title{ Multi-critical Behavior in Topological Phase Transitions }

\author{ S. \surname{Rufo} }
\email{srufo@cbpf.br}
\author{ Nei \surname{Lopes} }
\author{ Mucio A. \surname{Continentino} }
\affiliation{Centro Brasileiro de Pesquisas F\'isicas, \\Rua Dr. Xavier Sigaud, 150 - Urca, 22290-180,  Rio de Janeiro, RJ, Brazil}
\author{ M. A. R. \surname{Griffith}}
\affiliation{Universidade Federal de S\~{a}o Jo\~{a}o del-Rei, \\Rua , 22290-180,   S\~{a}o Jo\~{a}o del-Rei, MG, Brazil}

\date{\today }

\begin{abstract}
Topological phase transitions can be described by the theory of critical phenomena and identified by critical exponents  that define their universality classes. This is a consequence of the existence of a diverging length at the transition that has been identified as the penetration depth of the surface modes in the non-trivial topological phase.  In this paper, we characterize different universality classes of  topological transitions by determining their correlation length exponents directly from numerical calculations of the penetration length of the edge modes as a function of the distance to the topological transition. We consider generalizations of the topological non-trivial Su-Schriefer-Heeger (SSH) model, for the case of next next nearest neighbors hopping terms and in the presence of a synthetic potential. The latter allows the system to transit between two  universality classes with different correlation length  and dynamic critical exponents. It presents a line of multi-critical point in its phase diagram since the behavior of the Berry connection depends on the path it is approached. We compare our results with those obtained from a scaling approach to the Berry connection.

\end{abstract}

\maketitle

\section{Introduction}

Landau's paradigm for phase transitions constitutes one of the most successful theories of  modern physics. The main idea of this approach is to identify the symmetry breaking (SB) related to an ordered phase, such that an order parameter, that vanishes at the critical point can be defined~\cite{Landau1,ShunLivro}. This theory can be used in both classical and quantum phase transitions, and has been widely applied to predict the phase diagram of several systems, such as phases of water and magnetic phases~\cite{Landau2}.

On the other hand, there is a new different class of phase transitions whereupon this paradigm is no longer suitable. This class of phase transitions, which separates the phases of matter with different electronic Bloch states topology~\cite{ShunLivro,MucioLivro1,Alicea,Hasan,Sato,Fradkin} is called topological phase transitions (TPTs).

In distinction to Landau's approach, we are not able to define an order parameter related to the transition from the topological trivial to non-trivial phases. In general, there is no SB associated with a system that undergoes a TPT. Although  Landau's theory  is not useful to describe topological transitions, it is still possible to apply the theory of critical phenomena in order to identify critical exponents, scaling relations and finite size effects~\cite{GriMucio,MucioLivro2,Fadi,Cristiane,Cristiane2}.

TPTs that occur at zero temperature  are quantum phase transitions and  close enough to the quantum critical point (QCP), one can define both spacial and temporal characteristic {\it lengths} that diverge at the transition point, with critical exponents $\nu$ and $z$, respectively~\cite{MucioLivro1,StanleyLivro,Mucio3}. This divergence guarantees that at the QCP the whole system is correlated, which favors global, long range quantum fluctuations.

Instead of an order parameter, topological phases are characterized by topological invariants ($\mathcal{W}$), as the winding number~\cite{ShunLivro,Kosterlitz}. Unfortunately, this quantity is quantized, which makes it inappropriate  to be taken as an order parameter in a Landau approach.

Despite  these difficulties, several approaches have been used to obtain the universality classes of topological transitions. These include a direct determination of the correlation length exponent from numerical studies of the penetration length of the surface modes as a function of the distance to the transition~\cite{GriMucio,MucioLivro2}. This is the approach we adopt here and that allows us to identify different universality classes for topological transitions in isotropic systems.

Another interesting approach~\cite{Chen1,Chen3,Chen4,Chen2} relies on the idea that a topological invariant is a kind of correlation function and contains information about the correlation length of the system close to a topological transition. The Berry connection is found to obey a scaling form that allows to obtain the correlation length exponent.

For the one dimensional systems considered here, the energy dispersion close to the TPT can be written as,
\begin{equation}\label{Gap}
 E_k=\sqrt{ |g|^{2 \nu z} + k^{2 z}},
\end{equation}
where, $g=t-t_c$ is the distance to the QCP, at $t=t_c$~\cite{MucioLivro1,GriMucio}. At $k=0$, the gap for excitations, $\Delta=|g|^{\nu z}$ defines the gap exponent $\nu z $ where $\nu$ is the correlation length critical exponent, $z$ is the dynamical critical exponent, since at $g=0$, $E_k \propto k^z$. The identification of these exponents is a direct consequence that close to the quantum phase transition, the ground state energy density has the general scaling form, $\mathcal{E} \propto \sum_k E_k \propto |g|^{\nu(d+z)}$ where $d$ is the dimensionality of the system~\cite{MucioLivro2}.

In this paper to study the critical behavior near TPTs, we consider generalizations of the Su-Schrieffer-Heeger (SSH)~\cite{SSH} model. These models present complex phase diagrams, but still preserve some simplicity that makes them amenable to both analytical and numerical approaches. A direct numerical calculation of the penetration depth of the edge modes as a function of the distance to the topological transition yields the correlation length exponent $\nu$. This is compared with the results  from other approaches. In particular with that  based on the scaling properties of the  Berry connection~\cite{Chen1,Chen3,Chen4,Chen2}.

We show that the SSH model in the presence of a synthetic potential presents topological transitions that are in different universality classes as characterized by distinct values of their correlation length and dynamic critical exponents. It exhibits  in its phase diagram a line of multi-critical points whereupon the behavior of the Berry connection depends on the path it is approached.

The paper is organized as follows: in Section~\ref{SSHmodel}, we introduce the SSH model with next next nearest neighbors hopping, pointing out the new features and presenting its phase diagrams for different values of the parameters. In Section~\ref{Determination}, we calculate the correlation length critical exponents for the different TPTs using the penetration depth. We compare them to the results expected from the Berry connection approach. The TPTs of our model turn out to be in the same universality class characterized by dynamic exponent $z=1$ and correlation length exponent $\nu=1$.  We find perfect agreement between our direct numerical calculation of  the penetration depth of the edge states and the scaling behavior of the Berry connection. We confirm, as pointed out in Ref.~\cite{Chen2}, that in general there is no relation between the correlation length exponents and the {\it jumps} of the topological invariant across the TPTs.
Next, in Section~\ref{Breakdown}, we propose a modified SSH model in the presence of a synthetic potential that for a certain range of parameters gives rise to a quantum topological transition in the Lifshitz universality class with $z=2$ and $\nu=1/2$~\cite{Volovik,Volovik2,Imada}. Finally, in Section~\ref{Conc} we summarize the main results.

\section{The SSH model with long-range hopping terms }\label{SSHmodel}

\begin{figure}[h!]
  \centering
  \includegraphics[width=1\columnwidth]{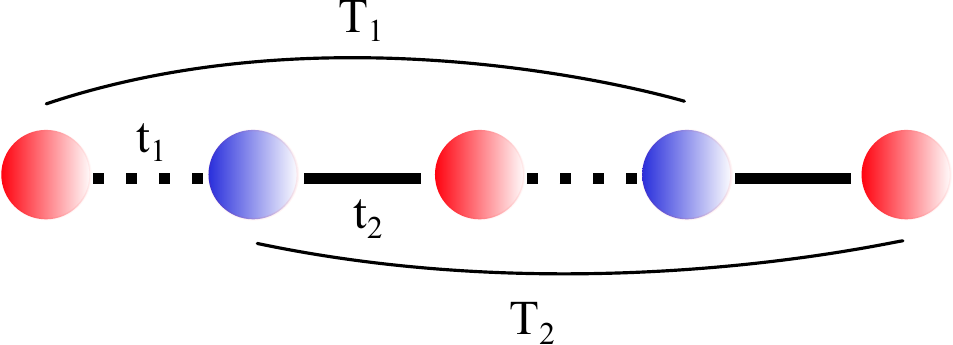}
  \caption{(Color online) The SSH model with next next nearest neighbors hopping terms. Each unit cell contains a pair of sub-lattices A and B, the red and blue spheres, respectively. The hopping terms connect different sub-lattices, $t_{1}$ within the unit cell while $t_{2}$, $T_{1}$ and $T_{2}$ out of the unit cell, as indicated.}\label{SSHFig}
\end{figure}

The nearest neighbor Su-Schrieffer-Heeger (SSH) model~\cite{SSH} was initially proposed for understanding the electronic properties of an organic polymer known as polyacetylene~\cite{ShunLivro}. This molecule is composed by a dimerized chain of Carbon (C) and Hydrogen (H), which can be physically described by a simple tight-binding Hamiltonian: a one-dimensional lattice of spinless fermions,  with staggered hopping amplitudes $t_1$ (intracell) and $t_2$ (intercell) that connect only atoms of Carbon in different sub-lattices of the unit cell, see Fig.~\ref{SSHFig} for additional hoppings $T_{1}=T_{2}=0$. This type of hopping structure only between different sub-lattices ensures the chiral symmetry of the system~\cite{Schnyder}.

We can think of the SSH model in the presence of next next nearest neighbors hopping (NNNN-hopping) terms $T_{1}$ and $T_{2}$ as a natural evolution of the original one~\cite{Linhu}, such that the Hamiltonian in real space is given by,
\begin{eqnarray}\label{SSH1}
 \mathcal{H}&=& t_1 \sum_i (c^{\dagger}_{A,i} c_{B,i} + h.c.) + t_2 \sum_i ( c^{\dagger}_{B,i} c_{A,i+1} + h.c.) \\
&+&T_1 \sum_i (c^{\dagger}_{A,i} c_{B,i+1} + h.c.) + T_2 \sum_i ( c^{\dagger}_{B,i} c_{A,i+2} + h.c.) \nonumber
\end{eqnarray}
where, the hopping terms $t_1$, $t_2$, $T_1$ and $T_2$ follow the scheme of Fig.~\ref{SSHFig}. In this equation, $c_{A,i}^{\dagger}$($c_{A,i}$) and $c_{B,i}^{\dagger}$($c_{B,i}$) are the creation (annihilation) fermionic operators that act in the sub-lattices A (red) and B (blue), respectively. After a Fourier transform, the Hamiltonian in $k$-space can be written as
\begin{equation}\label{SSHDiracEq}
\mathcal{H}_{k} = \sum_{i} h_{i}(k)\cdot \sigma_{i},
\end{equation}
the $\sigma_{i}$ are the Pauli matrices and
\begin{eqnarray}\label{h1h2h3}
  h_{1}(k)&=&t_{1}+t_{2}\cos(k)+T_{1}\cos(k)+T_{2}\cos(2k) \\ \nonumber
  h_{2}(k)&=&t_{2}\sin(k)-T_{1}\sin(k)+T_{2}\sin(2k) \\ \nonumber
  h_{3}(k)&=&0.
\end{eqnarray}
Notice, in special, that $h_{3}(k)=0$ ensures a chiral symmetry $\Pi=\sigma_{3}$. Furthermore, the model also possesses a time-reversal symmetry $\Theta=\mathcal{K}$ and an induced particle-hole symmetry $\Xi=\Pi\Theta$. Hence, the suitable symmetry class is AIII for the 1D model described by the hopping scheme of Fig.~\ref{SSHFig}, as discussed in~\cite{Chen2}.

In order to obtain the phase diagram of the system shown in  Fig.~\ref{SSHFig}, we have calculated the topological invariant winding number ($\mathcal{W}$) given by
\begin{equation}\label{winding}
  \mathcal{W}=\frac{1}{4 \pi i} \int_{0}^{2\pi} dk Tr ( \sigma_3 \mathcal{H}_k^{-1} \partial_k \mathcal{H}_k),
\end{equation}
where, $\sigma_3$ is the chiral operator and $\mathcal{H}_k$ is the kernel of the Hamiltonian in $k$-space.

\begin{figure}[t!]
\centering
\includegraphics[width=1\columnwidth]{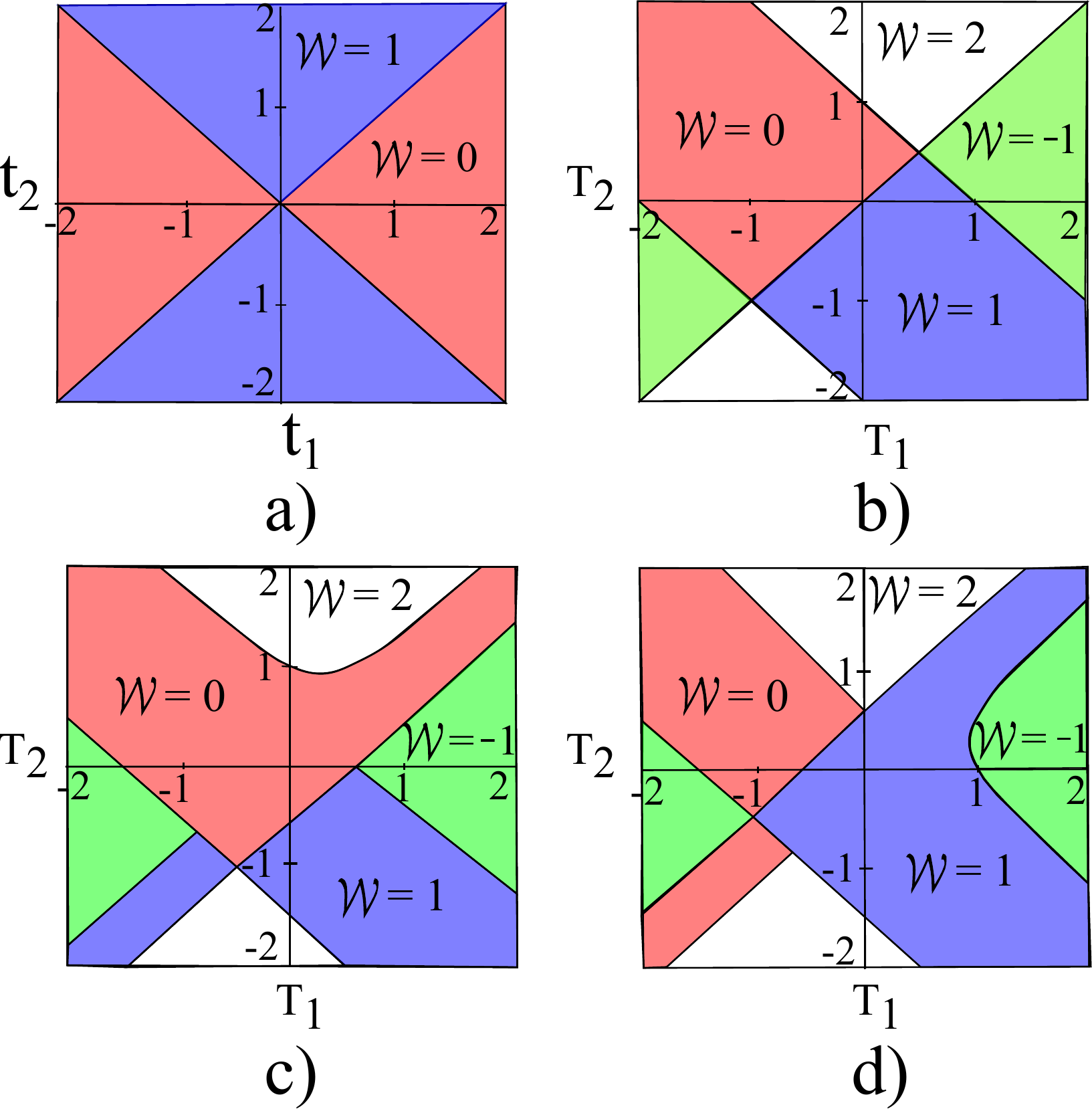}
\caption{(Color online) Phase diagrams of the SSH model. a) In the presence of only nearest neighbors hopping, $t_1$ and $t_2$. In b), c) and d) the NNNN-hopping terms $T_{1}$ and $T_{2}$ are included for several fixed  values of $t_{1}$ and $t_{2}$. Namely, $t_{1}=t_{2}=1$, $t_{1}=2t_{2}=1$ and $2t_{1}=t_{2}=1$ for b), c) and d), respectively. The topological invariants denote the trivial $\mathcal{W}=0$ (red) and non-trivial $\mathcal{W}\neq0$ topological phases. The long-range hopping terms give rise to two new topological phases with $\mathcal{W}=-1$ (green) and $\mathcal{W}=2$ (white), in addition to the $\mathcal{W}=1$ (purple) and $\mathcal{W}=0$ trivial phase obtained in a).} \label{Phase}
\end{figure}

It is well-known that this model for $T_1=T_2=0$, presents a TPT for $|t_1|=|t_2|$, separating two  phases with distinct topological properties. The phase diagram is presented in Fig.~\ref{Phase}~a). The trivial and non-trivial phases are identified by the invariant values $\mathcal{W}=0$ and $\mathcal{W}=1$, for the respective intervals $|t_1|>|t_2|$ and $|t_2|>|t_1|$ along the diagonal. The topologically non-trivial phase is protected by the chiral symmetry.

The addition of the NNNN-hopping terms $T_{1}$ and $T_{2}$ between the different sub-lattices preserves the chiral symmetry and provides a very rich phase diagram containing four topological phases of matter, i.e., two new phases further than in the nearest neighbors case. These phase diagrams are presented in Figs.~\ref{Phase}~b), c) and d), for fixed values of the nearest neighbors hopping. The different topological phases  are classified by $\mathcal{W}=-1,\,0,\,1$ and~$2$.

As we have pointed out before, the non-existence of an order parameter and its correlation function for topological transitions is a serious challenge to define a correlation length for these transitions.  However, an essential feature of non-trivial topological phases is the existence of gapless localized surface states. For the SSH chain in the non-trivial phases, these zero energy modes are mostly localized at the edges of the chain.
However, as the system approaches a topological phase transition, these modes delocalize and penetrate into the bulk of the chain. This occurs exponentially with a decay characterized by a length  $\xi$ that depends on the distance to the topological transition as $\xi=|g|^{-\nu}$. We refer to $\xi$ as the correlation length and $\nu$ the correlation length critical exponent ~\cite{Zhou,Chu}.
Here, we obtain this exponent numerically, as shown below. We also compare it with the value expected from the scaling properties of the Berry connection, as proposed in Refs.~\cite{Chen1, Chen2}.
Without loss of generality, we consider $t_{1}=t_{2}=1$ in Fig.~\ref{Phase}~b), since taking them differently does not increase the number of topological phases.

\section{ Determination of the critical exponent $\nu$ }\label{Determination}

In this section, we obtain initially the correlation length critical exponent $\nu$ and the dynamic critical exponent $z$ of the SSH model with NNNN-hopping terms using an expansion of the energy dispersion relations close to gap-closing momenta. Next, we present the results from the penetration depth calculations and compare them with those obtained  from the scaling of the Berry connection approach.
\begin{table*}[ht!]
\centering
\begin{tabular}{|p{2,0cm}|l|l|l|l|l|l|l|l|l|l|}
\hline \hline
    $(T_1,T_2)$ &$ \Delta W$ &$(k_{1,0},k_{2,0})$ & $A_1$& $A_2$ & $A_3$ & $A_4$ & $E_k$& $\Delta \rightarrow 0 $ &z & $\nu$  \\
\hline \hline
    $(0.0, 1.0)$   &2 &$(\frac{2 \pi}{3},-)$ & (0,-)& (3.0,-) & (-1.732,-) & (-0.75,-) & $\sqrt{ A_2 }k$ & $(T_1+T_2-1)^2$ &1 & 1 \\
    $(1.5,-0.5)$   &2 &$(\frac{2 \pi}{3},-)$ & (0,-)& (9.75,-) & (-1.732.-) & (-5.812,-)& $\sqrt{A_2} k$& $(T_1+T_2-1)^2$&1 & 1 \\
    $(-1.5, -1.5)$ &3 &$(\pi,0.58^{*})$    & (0,0)& (30.25,10.083) & (0,12.161) & (-19.021,-4.507) &  $\sqrt{ A_2 }k$& $(T_1-T_2)^2$ &1 & 1 \\
    $(0.0, 0.0)$   &1 &$(\pi,-)$             & (0,-)& (1,-) & (0,-) & (-0.083,-)& $\sqrt{A_2 }k$& $(T_1-T_2)^2$ & 1& 1  \\
    $(1.0, 1.0)$   &3 &$(\pi$ , $\frac{\pi}{2})$  & (0,0)& (4.0,8.0) & (0,-8.0) & (-4.333,-2.667) & $\sqrt{ A_2} k$& $(T_1-T_2)^2$ &1 &1  \\
    $(-1.5,-0.5)$  &1 &$(0,-)$ & (0,-)&  (2.25,-) & (0,-) & (2.312,-)& $\sqrt{A_2} k$& $(T_1+T_2+2)^2$& 1 & 1\\
    $(-0.5,-1.5)$  &1 &$(0,-)$ & (0,-)&  (2,25,-) & (0,-) & (2.312,-)& $\sqrt{ A_2} k$& $(T_1+T_2+2)^2$ & 1 & 1\\
    $(0.33,0.33)$  &1 &$(\pi,-)$  & (0,-)& (0.0001,-) & (0,-) & (0.007,-)& $\sqrt{A_2 }k$& $(T_1+T_2-1)^2$& 1 & 1\\
\hline \hline
\end{tabular}
\caption{Critical points and their critical exponents for the SSH model with NNNN-hopping terms obtained by an analysis of the energy dispersions. After an expansion around the gap-closing points $k_{1,0}$ and $k_{2,0}$, as given in Eq.~(\ref{Aproximation}), we present the coefficients $A_{n}$, with $n=1\cdots 4$ and the gap functions $\Delta$. The product $A_{n}k^{n}$ determines the dominant behavior of the energy $E_{k}$.\\
*-the complete value is $0.5856855435$.} \label{table1}
\end{table*}

\subsection{Energy dispersion }\label{EDispersion}

We can obtain the dispersion relations of the excitations in the chain close to the critical lines  of the phase diagram presented in Fig.~\ref{Phase}~b). We  diagonalize the Hamiltonian, Eq.~(\ref{SSHDiracEq}), to write the energy dispersions as
\begin{equation}\label{Dispersion}
  E_k = \pm \sqrt{a + b \cos(k) + c \cos^{2}(k) + d \cos^{3}(k) },
\end{equation}
where $a=[(t_{1}-T_{2})^2+(t_{2}-T_{1})^2]$, $b=2(t_{1}T_{1}+t_{2}T_{2}+t_{1}t_{2})-6T_1T_2$, $c=4(t_1T_{2}+t_2T_{1})$ and $d=8T_1T_2$.

\begin{figure}[t!]
\centering
\includegraphics[width=1\columnwidth]{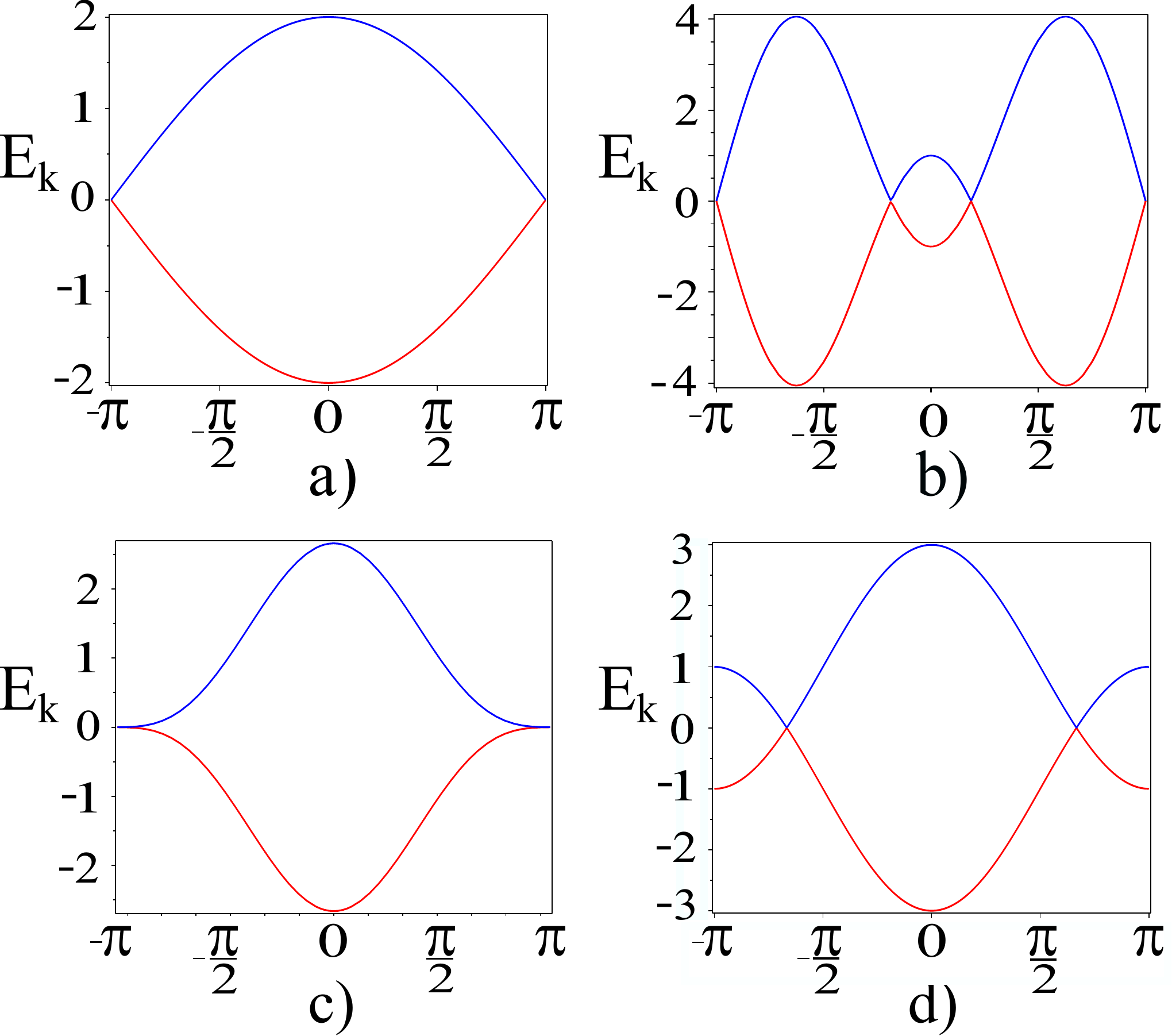}
\caption{(Color online) The energy dispersions corresponding to the QCPs  of Fig.~\ref{Phase}~b) for $t_1 =t_2=1$. We investigate the dispersions close to the critical points ($T_1,T_2$) and gap-closing momenta $k_{0}$ as follows: a) (0,0) for $k_{0}=\pi$, b) (-1.5,-1.5) for $k_{0}=\pi,\,0.58$, c) (0.33,0.33) for $k_{0}=\pi$ and d) (0,1) for $k_{0}=2\pi/3$.}\label{Dispersions}
\end{figure}

Fig.~\ref{Dispersions} shows $E_k$ as a function of the momentum $k$ at four QCPs $(T_1,T_2)$ in the phase diagram of Fig.~\ref{Phase}~b). Once we fix $T_{1}$ for each transition,  the jump of the invariant is given by $\Delta W= |W(T_2>T_{2c})-W(T_2<T_{2c})|$. In Fig.~\ref{Dispersions}~a) and c) we notice trivial to non-trivial phase transitions with $\Delta\mathcal{W}=1$. In Fig.~\ref{Dispersions}~b), the QCP separates two topologically non-trivial phases with $\Delta\mathcal{W}=3$. For Fig.~\ref{Dispersions}~d), the transition occurs from a non-trivial to trivial phase with $\Delta\mathcal{W}=2$.

So, the study of the present model reveals three different values for the jumps of the topological invariant, i.e., $\Delta\mathcal{W}=1,\,2$ and $3$. According to the relation $\nu = \frac{1}{\Delta \mathcal{W}}$~\cite{Chen1}, we expect to obtain different values for the correlation length exponents and consequently different universality classes for these transitions, besides the case  $\Delta\mathcal{W}=1$ for the trivial to non-trivial transitions. The validity of this relation will be discussed below, when we present our results for the correlation length exponents. We have also searched for some clear relation between the gapless points in the Brillouin Zone and $\Delta\mathcal{W}$. For this purpose, we tested all points explored in Table I and concluded that  $\Delta\mathcal{W}$ cannot account for the number of gapless points across the TPT. For instance, for the first point of Table I ($T_{1},T_{2})=(0,1)$  the number of gapless points (G) and  $\Delta\mathcal{W}$ coincide  $\Delta\mathcal{W}=G=2$. Nevertheless, for the 3rd point of the Table I ($-1.5,-1.5$) the relation is given by $G=\Delta\mathcal{W}+1=4$.

In a first view, all behaviors around the gap-closing momenta $k_{0}$ in Fig.~\ref{Dispersions} seem to be linear, except in Fig.~\ref{Dispersions}~c). To verify this, we look more closely, with an expansion around $k_{0}$ and inspect the critical exponents $\nu$ and $z$ of  Eq.~(\ref{Gap}).
It is also important to analyze the dominant terms of  Eq.~(\ref{Dispersion}) at the vicinity of these QCPs. For this reason, we have expanded the energy dispersions around the gap-closing momenta $k_{0}$. The general form of the expansion is given by
\begin{eqnarray}\label{Aproximation}
  E_k &=& \pm \sqrt{ |g|^{2\nu z} + \sum^{4}_{n=1} A_n  k^{n} },
\end{eqnarray}
such that at the QCP, $k=k_0$ and the gap function $\Delta = |g|^{\nu z}$ should go to zero. Near to the QCP, the dominant $k^n$ can be identified, which implies that  Eq.~(\ref{Aproximation}) may be written as
\begin{eqnarray}\label{Aproximation2}
  E_{k} \propto k^{\frac{n}{2}} = k^z,
\end{eqnarray}
which shows that at the QCP ($g=0$)  the shape of the spectra at the the gap-closing points is dominated by $k^z$.  We perform this analysis for the points presented in Fig.~\ref{Dispersions}, as well as, for other points of the phase diagrams Fig.~\ref{Phase}~b), for which the results are summarized in Table~\ref{table1}. This also includes all $A_{n}$ coefficients, to make clear the more intense and dominant ones together with the appropriate $k^{n}$, besides the $E_{k}$ and gap behavior and its critical exponents $z$ and $\nu$.

We call attention to the point $(-1.5,-1.5)$ in Table~\ref{table1}, which possesses two values of gap-closing momenta. Once again, for the same power of $k$,  the more intense coefficient $A_{i}$ will determine the relevant $k_{0}$ for each TPT. In this case, the coefficient $A_{2}(k_{0}=\pi)\simeq 30$ is larger than $A_{2}(k_{0}=0.58*)\simeq10$, hence the gap-closing dominant value is $k_{0}=\pi$. The same statement is valid for other points.

Thus, from Table~\ref{table1} we notice that  the dynamic critical exponents $z$ takes the value of unity, i.e., $z=1$, due to the linear behavior of the energy $E_{k}$ close to $k_0$, for all  TPTs  investigated. This includes the case of Fig.~\ref{Dispersions}~c). Since the gap vanishes linearly at the gap-closing moments, once the dynamical critical exponent is identified,  the critical exponent $\nu=1$ can also  be obtained from the gap equation.

\subsection{Penetration Depth }\label{PDepth}

\begin{figure}[t!]
\centering
\includegraphics[width=1\columnwidth]{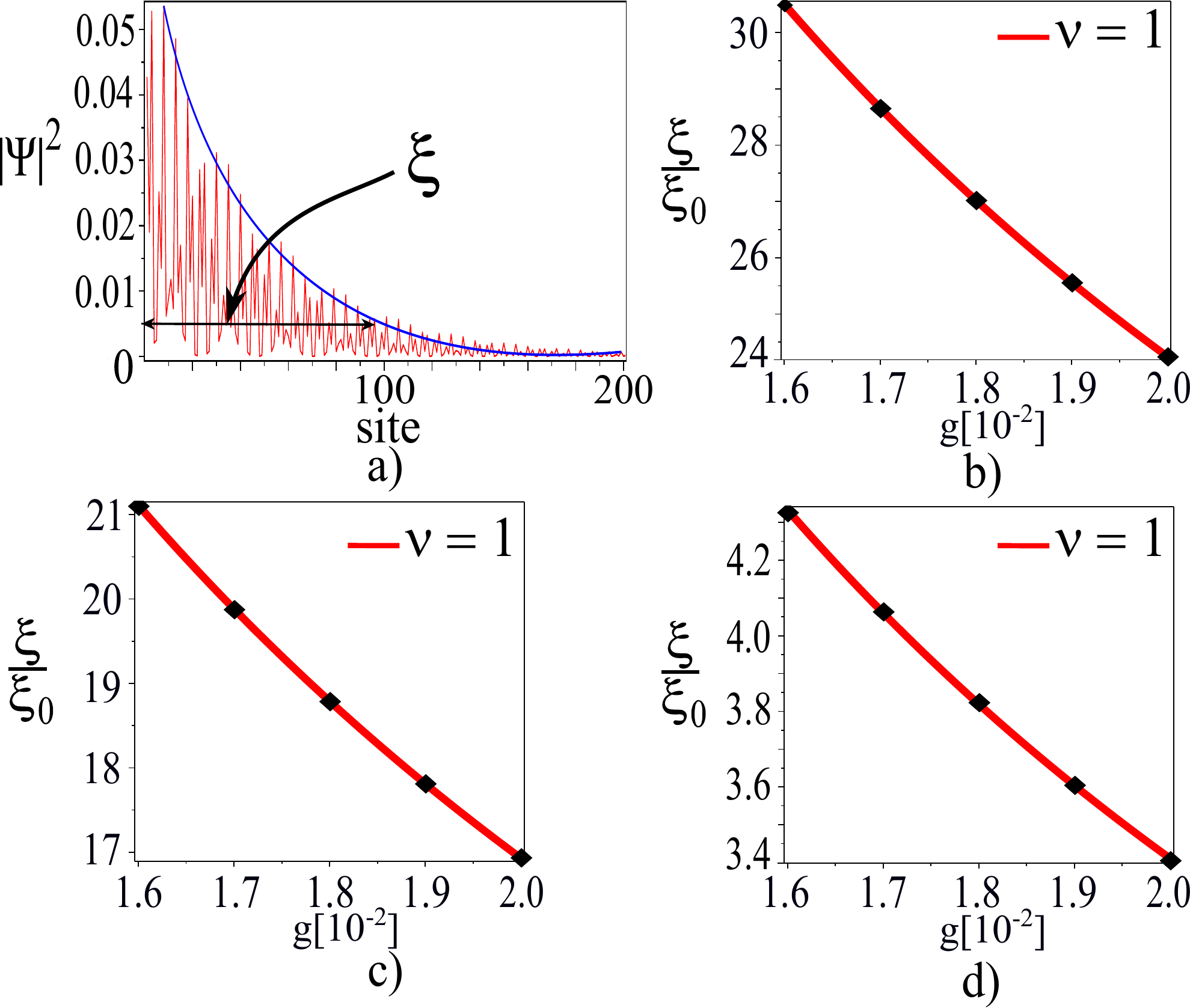}
\caption{(Color online) The correspondence between the penetration depth $\zeta$ and the characteristic length $\xi$ is presented in a), close to the critical point $(T_{1},T_{2})=(-1.5,-1,5)$ where the amplitude of the wave function is observed to decay exponentially into the bulk (thin continuous blue line).  In b), ~c) and d) we  show plots of $\zeta=\xi$ (black squares) as a function of the distance  $g$ to the QCP. These are fitted by $\xi =\xi_0 |g|^{-\nu}$, from which we extract the correlation length exponents $\nu$. The critical points explored are $(0,0)$, $(0,1)$ and $(-1.5,-1.5)$ in b),~c) and d), respectively. The fitting is very satisfactory for all cases and agrees with $\nu=1$, independent of the jump of the invariant $\Delta\mathcal{W}$ in each case. The results presented are for a chain with $N = 1600$ sites.}
\label{Depth}
\end{figure}

In general, at a phase transition  there is only one diverging length, the correlation length $\xi$ that dominates the transition near to the QCP. Accordingly, as shown below, it is possible to identify the diverging penetration depth of the edge modes  as the correlation length.

Since the amplitude of the wave function of the zero energy edge mode is observed to decay exponentially in the bulk, as shown in Fig.~\ref{Depth}~a) (continuous line), i.e., $\sqrt{|\Psi(x)|^2}=\sqrt{|\Psi(0)|^2}e^{-x/\xi}$, the penetration depth $\zeta$ is easily obtained as the distance between the edge of the chain and the point inside the chain at which this amplitude has decreased to $e^{-1}$ from its value at the edge.  In  mathematical terms,  $\sqrt{|\Psi(\zeta)|^2}=\sqrt{|\Psi(0)|^2}/e$. For simplicity, we introduced above a continuous variable $x=(n-1)a$ where $n$ is the site index and $a$ the average atomic distance. Here we identify the penetration depth with the correlation length, i.e., $\xi = \zeta$ as shown in Fig.~\ref{Depth}~a).
Notice that in this figure,  the amplitude of the surface mode wave function is reduced to $e^{-1}$ of its value at the surface, at approximately the $100$-th site. The choice of this factor ($e^{-1}$) is for convenience and does not affect the main results.

After the diagonalization of the Hamiltonian, Eq.~(\ref{SSH1}) in real space, we obtain for a chain with N sites, $N$ energies and its $N$ eigenstates. In order to verify whether it makes sense to identify the penetration depth as the characteristic diverging length, we obtain these lengths for several distances $g$ to the critical point. The results for $\xi$ as function of $|g|$ are presented in Fig.~\ref{Depth}~b),~c) and d) as black points.  The continuous (red) lines represent fittings of the penetration depth following  $\xi = \xi_0 \left|g \right|^{-\nu}$ and allow to obtain the correlation length exponents that turn out to be $\nu=1$.
Investigating the localization of the edges states in the topological phases with $\mathcal{W}=1$,$-1$ and $2$ from Fig.~\ref{Phase} b), we verified that at the left end of the chain the zero-energy edges states are localized only at  sub-lattice A, while at the right end, the localization emerges only at sub-lattice B.
We also observe that the jump of the topological invariant $\Delta\mathcal{W}$ does not have any relation with the critical exponents $\nu$. The reason for that will be discussed further on in the text.
\begin{table}[t!]
\begin{tabular}{|p{2,0cm}|l|l|}
\hline \hline
    $(T_1,T_2)$ &$\Delta \mathcal{W}$ &$\nu$ \\
\hline \hline
    $(-0.5,1.5)$   &$2$	 & $\nu_v=0.998101 \ - \ \nu_h=1.011832$ \\

    $(0,1)$        &$2$  & $\nu_v=0.980898 \ - \ \nu_h=0.990579$ \\

    $(1.5,-0.5)$   &$2_{(-1\rightarrow1)}$  &$\nu_v=1.008451 \ - \ \nu_h=0.994491$ \\

    $(1.5,-0.5)$   &$2_{(1\rightarrow-1)}$  &$\nu_v=0.993665 \ - \ \nu_h=1.006287$  \\

    $(-1.5,-1.5)$  &$3_{(2\rightarrow-1)}$  &$\nu_v=1.068392 \ - \ \nu_h=1.032452$ \\

    $(-1.5,-1.5)$  &$3_{(-1\rightarrow2)}$  &$\nu_v=0.976282 \ - \ \nu_h=1.098094$ \\

    $(0,0)$        &$1$  &$\nu_v=0.986789 \ - \ \nu_h=1.009757$ \\

    $(1,1)$        &$3_{(2\rightarrow-1)}$  &$\nu_v=1.003701 \ - \ \nu_h=1.008982$ \\

    $(1,1)$        &$3_{(-1\rightarrow2)}$  &$\nu_v=1.009280 \ - \  \nu_h=0.993825$ \\

    $(-1.5,-0.5)$  &$1$  &$\nu_v=1.058208 \ - \ \nu_h=1.022701$ \\

    $(-0.5,-1.5)$  &$1$  &$\nu_v=1.011765 \ - \ \nu_h=1.053610$ \\
\hline \hline
\end{tabular}
\caption{Critical points and their critical exponents $\nu$ for the SSH model with NNNN-hopping terms obtained by the penetration depth. The jump of the topological invariant $\Delta\mathcal{W}$, $\nu_h$ (horizontal approach) and $\nu_v$ (vertical approach) are presented to evidence how the path to the critical point affects the critical exponent $\nu$. We observe that all $\nu$ values are very close to unit. A couple of TPT present the same $\Delta\mathcal{W}$ value, for these cases an index was included to indicate the path.}
\label{tabledois}\end{table}

Following this approach, we extend the numerical study to verify how the way we approach to the QCP  affects the value of the critical exponent $\nu$. The extended results are presented in Table~\ref{tabledois}, for several values of critical points. In this table, we have cases with different paths for TPT, but the same $\Delta\mathcal{W}$. For these cases, we have included an index to specify the path to the transition.

If we look to the phase diagram of Fig.~\ref{Phase}, we have at least two different paths to approach a  TPT. One is to fix  $T_{1}$ and perform a vertical approach to the critical point as $g=|T_{2}-T_{2c}|$ goes to zero. The other  is to fix  $T_{2}$ and perform a horizontal approach to the critical point for $g=|T_{1}-T_{1c}|$. The critical exponents obtained by the vertical and horizontal approaches are called $\nu_v$ and $\nu_{h}$, respectively. The results of  Table~\ref{tabledois} show that it does not matter the path of approach to the TPT adopted. The same jump of the topological invariant is observed for different paths or  distinct approaches to the critical point. All the critical exponents are very close to  unit and do not seem to be affected by the choice of the path.

In summary, the values for the correlation length exponents $\nu$ for  the SSH model with NNNN-hopping terms obtained from a numerical calculation of the penetration depth  (see Fig.~\ref{Depth} and Table~\ref{tabledois}) are in excellent agreement with those obtained from an analysis of the dispersion relations at the gap-closing points  given in Table~\ref{table1} .

\subsection{Berry connection}\label{Berry}

Recently, Chen et al.~\cite{Chen1,Chen2} proposed a different method to obtain the correlation length of topological transitions. It is based on the idea that a topological invariant may be considered as a kind of correlation function and can be used to extract a correlation length. In particular, in the non-trivial topological phase it contains information on the edge-states decay length.

A topological phase is usually characterized by some proper topological invariant $\mathcal{W}$. If $F(k,t)$ is the pertinent curvature function for the problem in $d$ dimensions, we can define a topological invariant as
\begin{equation}\label{TopInv}
\mathcal{W}=\frac{1}{(2\pi)^{d}}\int_{0}^{2\pi}F(k,t) d^dk
\end{equation}
where, $t$ indicates a tuning parameter that can drive a topological transition, in the present case it is related to hopping energy terms.

In particular, for $1D$ the curvature function is identified as the Berry connection~\cite{Haldane,Kane}. For the Dirac Hamiltonian $\mathcal{H}_{k}=\left(h_{1},h_{2},0\right)$ this can be written in terms of its elements as~\cite{Chen1}
\begin{equation}\label{BerryConnection}
F(k,t)=R \,\frac{h_{2}\partial_{k}h_{1}-h_{1}\partial_{k}h_{2}}{2\varepsilon^2},
\end{equation}
such that $\varepsilon(k)=\pm\sqrt{h_{1}^2+h_{2}^2}$ is the energy dispersion relation and $R$ is a renormalization term that ensures an integer value for the topological invariant, as expected.

The Berry connection is gauge dependent, so it is important to make a convenient choice for our purpose. The characteristic length scale $\xi$ of the system is known to diverge at the QCP~\cite{Mucio4,Evert}, and one simple way to identify it is to consider that the Berry connection has a Lorentzian shape close to the topological gap-closing transition~\cite{Chen1}. Chen et al.~\cite{Chen1} assume that this has the simple scaling form near to a gap-closing point $k_{0}$,
 \begin{equation}\label{BerryLor}
F(k,g)=\frac{F(k_{0},g)}{1 \pm \xi^2 {k}^{2}}.
 \end{equation}
with the correlation length,
\begin{equation}\label{ClengthBC}
\xi = \xi_0 \left|g \right|^{-\nu},
\end{equation}
where $g=t-t_{c}$ and the energy hopping term controls the TPT.

For a numerical analysis close to the QCP and in the vicinity of the gap-closing points, we consider the  transition lines of Fig.~\ref{Phase}~b). This phase diagram presents the distance to the critical points $g=T_{2}-T_{1}$, $T_{2}+T_{1}+2$ and $T_{2}+T_{1}-1$, with their respective gap-closing points $k_{0}=\pi,\,0$ and $2\pi/3$.  The expansion of the Hamiltonian terms $h_{1}$ and $h_{2}$, from Eq.~(\ref{h1h2h3}), around each of these $k_{0}$  yields
\begin{description}
  \item[$g=T_{2}-T_{1}$ and $k_{0}=\pi$]
\begin{eqnarray}\label{line1}
  h_{1}&=&(T_{2}-T_{1})-\frac{1}{2}(4T_{2}-T_{1}-1)\delta k^2 \nonumber \\
  h_{2}&=&(2T_{2}+T_{1}-1)\delta k
\end{eqnarray}
  \item[$g=T_{2}+T_{1}+2$ and $k_{0}=0$]
\begin{eqnarray}\label{line2}
  h_{1}&=&(T_{2}+T_{1}+2)-\frac{1}{2}(4T_{2}+T_{1}+1)\delta k^2 \nonumber \\
  h_{2}&=&(2T_{2}-T_{1}+1)\delta k
\end{eqnarray}
  \item[$g=T_{2}+T_{1}-1$ and $k_{0}=2\pi/3$]
\begin{eqnarray}\label{line3}
  h_{1}&=&-\frac{1}{2}(T_{2}+T_{1}-1)+\frac{\sqrt{3}}{2}(2T_{2}-T_{1}-1)\delta k  \nonumber \\
&+& \frac{1}{4}(4T_{2}+T_{1}+1)\delta k^2 \nonumber \\
  h_{2}&=&-\frac{\sqrt{3}}{2}(T_{2}+T_{1}-1)-\frac{1}{2}(2T_{2}-T_{1}+1)\delta k  \nonumber \\
&+& \frac{\sqrt{3}}{4}(4T_{2}+T_{1}-1)\delta k^2.
\end{eqnarray}
\end{description}
The first and second cases present the same behavior found previously for the SSH 1D-model with just nearest neighbors hopping, as discussed in Ref.~\cite{Chen1}, while the last one has a different expansion equation. Accordingly, for $k_{0}=\pi$ and $0$, we can rewrite the expansions in a more compact way as, $h_{1}=g t +B \delta k^2$ and $h_{2}=A \delta k$. Using  Eq.~(\ref{BerryConnection}), and after some calculations we obtain
\begin{eqnarray}\label{FSSH1sta}
  F(k, g) &=& \frac{R}{2}\frac{-Ag+AB\delta k^2}{g^2+(2Bg +A^2)\delta k^2} \nonumber  \\
   &=& \frac{\frac{R}{2}\frac{-Ag+AB\delta k^2}{g^2}}{1+[\frac{2Bg+A^2}{g^2}]\delta k^2}  \\
   &=& \frac{F(k_{0},g)}{1+\xi^2\delta k^2},  \nonumber
\end{eqnarray}
where $F(k_{0},\delta t)=-\frac{R}{2}Ag^{-1}$, in agreement with the scaling form of Chen et al.~\cite{Chen1}.  For the characteristic length scale $\xi$ we have
\begin{equation}\label{CharacScale1}
  \xi^2 = \left[\frac{2B}{g}+\frac{A^2}{g^2}\right],
\end{equation}
since the second term diverges more quickly, it becomes the dominant behavior close to the QCP. Hence,
\begin{eqnarray}
  \xi &=& \left[\frac{A^2}{g^2}\right]^{1/2} \propto | g |^{-1}\,\,\,\Rightarrow\nu=1.\label{Eq18}
\end{eqnarray}

Analogously, for the case with $k_{0}=2\pi/3$ we can write $h_{1}=ag+ A\delta k+ B\delta k^2$ and $h_{2}=a^{\prime}g+ A^{\prime}\delta k+ B^{\prime}\delta k^2$ to obtain for the Berry connection, Eq.~(\ref{BerryConnection}), close to the gap-closing point
\begin{equation}
  F(k, g) = \frac{F(k_{0},g)}{1+\left[ \frac{2(aB+a^{\prime}B^{\prime})g+(A^2+{A^{\prime}}^2)}{(a^2+{a^{\prime}}^2)g^2} \right]\delta k^2},
\end{equation}
with $F(k_{0},g)=\frac{R}{2}\left( \frac{a^{\prime}A-aA^{\prime}}{a^2+{a^{\prime}}^2} \right)g^{-1}$. Similar to Eq.~(\ref{CharacScale1}), the dominant behavior is
\begin{eqnarray}
  \xi &=& \left[\frac{A^2+{A^{\prime}}^2}{(a^2+{a^{\prime}}^2)g^2}\right]^{1/2} \propto | g |^{-1}\,\,\Rightarrow\nu=1.\label{Eq20}
\end{eqnarray}

\begin{figure}[t!]
\centering
  \includegraphics[width=1\columnwidth]{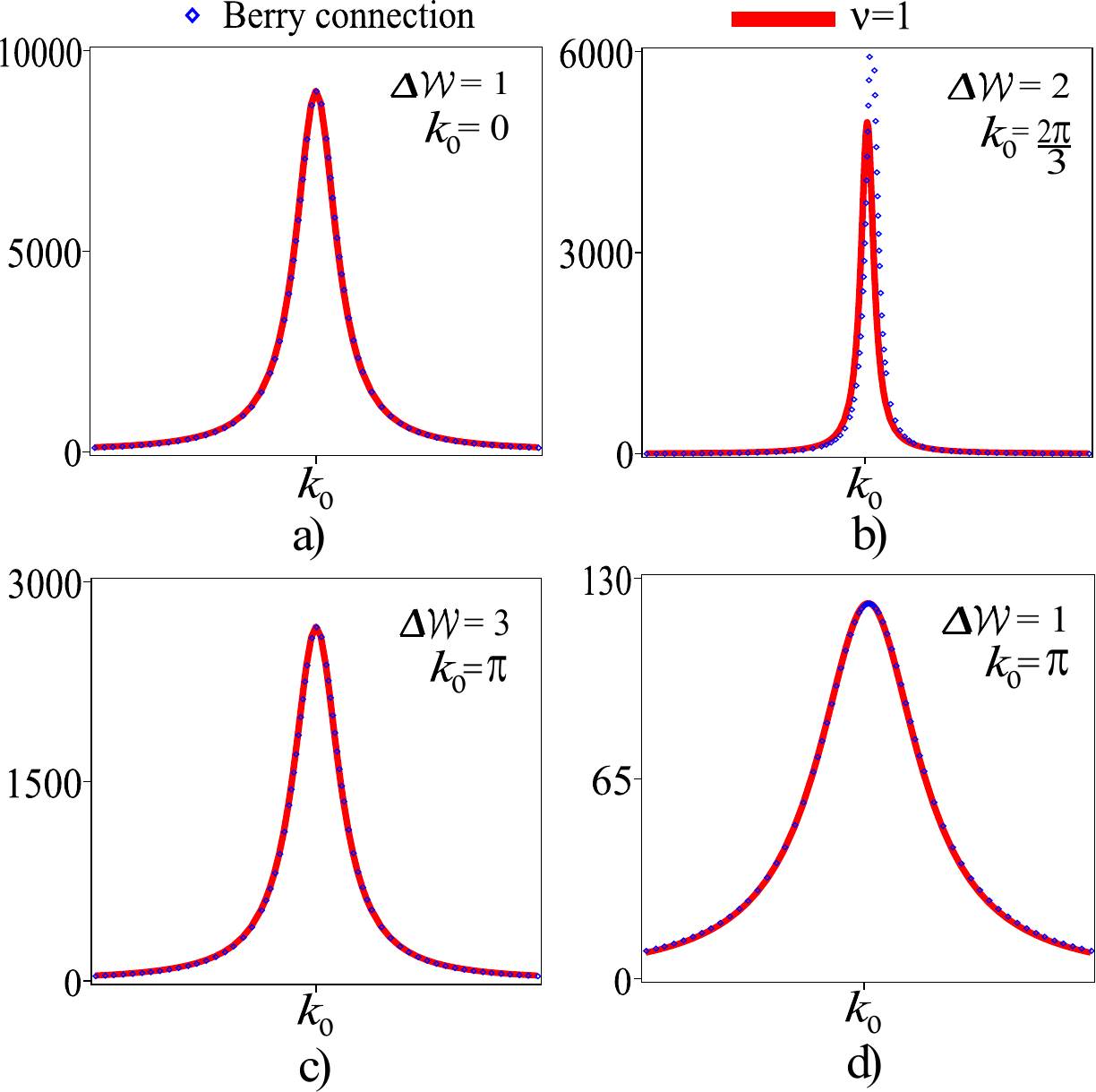}
  \caption{(Color online)  The fitting between the Berry (squares) and the Lorentzian (solid line) curvatures. The critical points $(T_{1},T_{2})$ investigated  are a) $(-1.5,-0.5)$, b) $(1,0)$, c) $(1,1)$ and d) $(0.33,0.33)$. These points contemplate $\Delta \mathcal{W}=1,\,2,\,3$ and $k_{0}=0,\,\pi,\,2\pi/3$ values of the SSH model. Only  case  b) does not show a good fit as concerns the amplitude. For a Lorentzian shape, the width of the function is given by $1/\xi$. The best fitting using  Eq.~(\ref{BerryLor})  leads to $\nu=1$ for all cases. The distance to the critical point is of the order $g\sim 10^{-4}$.}\label{ChenFit}
\end{figure}
Following the ideas of Ref.~\cite{Chen1,Chen2}, we perform a numerical fit of the Berry connection Eq.~(\ref{BerryConnection}), with a Lorentzian shape as in Eq.~(\ref{BerryLor}). In the fitting process, we use for the Lorentzian $\xi_{0}$ from Eq.~(\ref{Eq18}) for $k_{0}=0,\,\pi$, and from Eq.~(\ref{Eq20}) for $k_{0}=2\pi/3$. Then, we extract the value of $\nu$ that satisfies the fitting.

In Fig.~\ref{ChenFit}, we show the fitting process for four distinct points in the phase diagram of Fig.~\ref{Phase}~b). In order to include all $\Delta \mathcal{W}=1,\,2,\,3$ and $k_{0}=0,\,2\pi/3,\,\pi$ obtained for the present SSH model, we investigate the critical points $(T_{1},T_{2})=(-1.5,-0.5),\,(1,0),\,(1,1)$ and $(0.33,0.33)$, in Fig.~\ref{ChenFit}~a), b), c) and d), respectively. We observe that only the case of Fig.~\ref{ChenFit}~b), where $k_{0}=2\pi/3$, does not present a perfect fit of the Berry connection by the Lorentzian shape, as concerns its amplitude.

It is important to note that although the amplitudes do not fit perfectly, the width of the functions still preserve a satisfactory fitting. In special, the width of the Lorentzian  is inversely proportional to the characteristic length,  $\xi$. So, at least in principle, to determine $\xi$ the satisfactory fitting of the width is sufficient. The results of Fig.~\ref{ChenFit} return $\nu=1$ for all cases. This again implies that $\xi$ does not depend on the jump of the invariant and seems to be more affected by the gap-closing point, particularly its amplitude, as discussed below.

So, for $k_{0}=0,\,\pi$ the fittings are   in good agreement with all aspects of the scaling function.  For $2\pi/3$ just the width, which is related to $\xi$  is preserved in the fitting process by a Lorentzian shape.

\section{Breakdown of Lorentz invariance in topological phase transitions}\label{Breakdown}

The topological transitions studied so far are all Lorentz invariant, characterized by the dynamic exponent $z=1$. Furthermore, they present a correlation length $\nu=1$ as obtained by different methods.

With the purpose of investigating new universality classes and the success of the methods used above, we consider an extended SSH model with just nearest neighbors hopping, i.e., for $T_{1}=T_{2}=0$, but that includes a term $V=-x\sin(k) \sigma_2$ in the Hamiltonian, Eq.~(\ref{SSHDiracEq}).  Equivalently, this enters as a contribution to the term $h_{2}(k)$ of Eq.~(\ref{h1h2h3}), but with only nearest neighbors hopping. Above, $x$ works as an external control parameter for the synthetic potential $V$. In real space this term corresponds to an inter sub-lattice, antisymmetric hopping.
Notice that since the term $V$ is proportional to $\sigma_{2}$ it does not break the chiral symmetry. Accordingly, Eq.~(\ref{winding}) remains valid to obtain the topological invariant $\mathcal{W}$.

The spectrum of the system close to $k_0=\pi$ is given by,
\begin{equation}\label{quadratic}
  E_k = \sqrt{(t_1-t_2)^{2} + A_2(x) k^2 + A_4(x) k^4 },
\end{equation}
where, $A_{2}(x) = t_1t_2-2t_2 x+x^2$ and $A_{4}(x)=-\frac{t_{1}t_{2}}{12}+\frac{2t_{2}x}{3}-\frac{x^2}{3}$.

Notice that the gap closes as $g^{2\nu z}=(t_{1}-t_{2})^{2}$, which implies $\nu z=1$. For $A_{2} \ne 0$ the quadratic term always determines the critical behavior. However, there is a region where the quartic term $A_{4}k^{4}$ is much larger than $A_{2}k^{2}$. Then, even for small $k$ the quartic term may become dominant and we can observe a crossover from $z=1$ to $z=2$.
For $A_2=0$, the model is in a different universality class with $z=2$ and $\nu=1/2$. The situation where the quadratic term dominates leading to $\nu=1$ was already discussed in Sec.~\ref{EDispersion}.

\begin{figure}[t!]
  \centering
  \includegraphics[width=1\columnwidth]{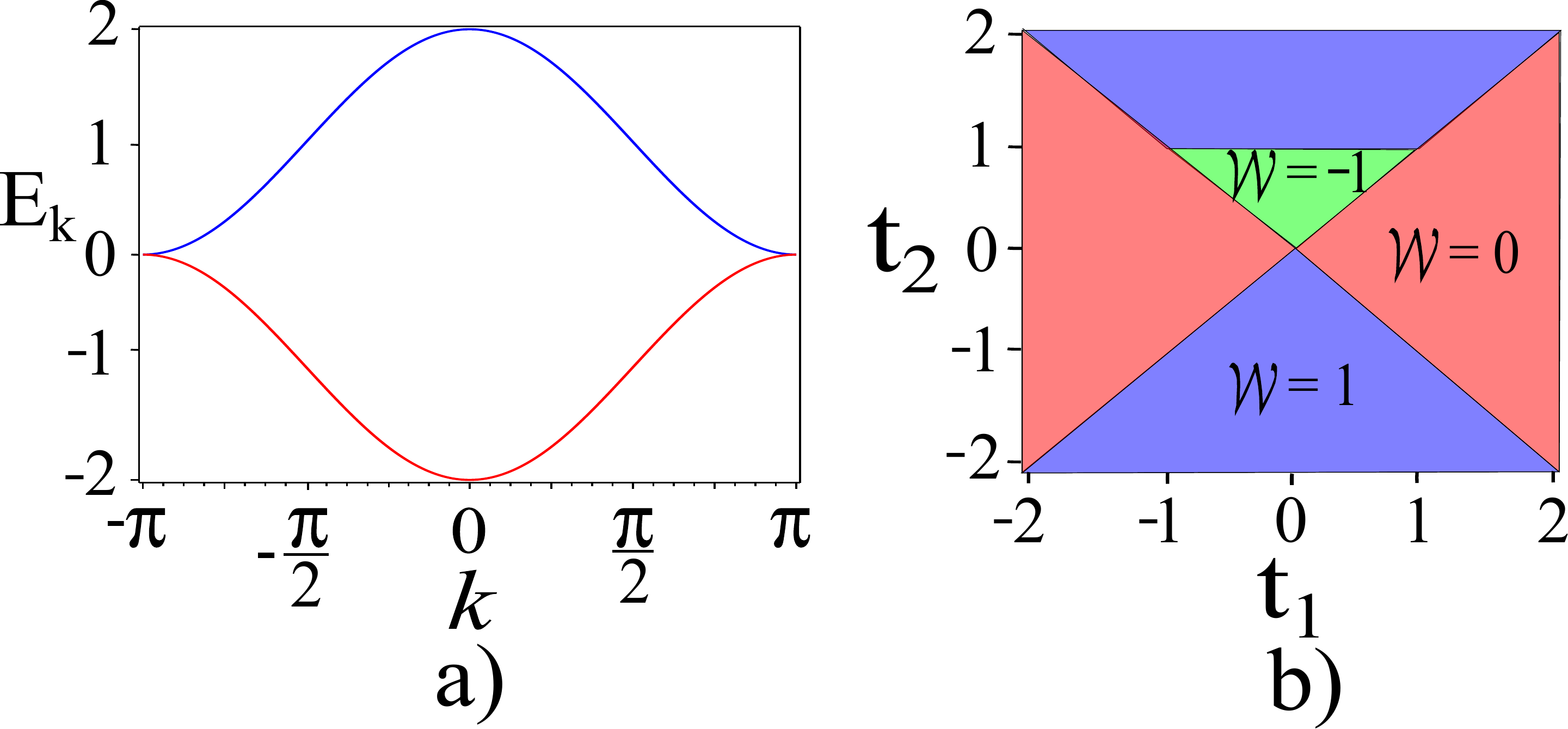}
  \caption{(Color online) a) Energy dispersions at the QCP for $x=1$, $t_1=1$ and $t_2=1$. The parabolic gap-closing between the eigenvalues  occur for $k=k_0=\pm \pi$. b) The phase diagram of the modified SSH model in the presence of a synthetic potential with  $x=1$. The topologically non-trivial phases correspond to $\mathcal{W}=-1$ (green), $\mathcal{W}=1$ (purple) and the topologically trivial phase with $\mathcal{W}=0$ (red). There are no edge states in the latter. In particular, for $t_1=t_2=1$ the energy dispersion is quadratic around this QCP.}\label{phaseX}
\end{figure}

In this section, we study the case $A_{2}=0$. This requires that
\begin{equation}\label{parameter}
 x^2 - 2t_2 x + t_1t_2 = 0,
\end{equation}
which together with the condition $E_{k_0}=0$ provide the criterion for a QCP with a quadratic dispersion. Notice that the line, $x=t_1=t_2$ in the phase diagram satisfies these conditions simultaneously. For instance, for fixed $x=1$, the point $t_{1}=t_{2}=1$ presents a quadratic gap-closing point at $k_0= \pm \pi$, see Fig.~\ref{phaseX}~a). In a more general way, we verify that, for $t_{2}=x$ and $x\geqslant1$ the system is always gapless with the presence of Dirac massless points.

To elucidate the nature of the TPT, we propose to study the phase diagram shown in Fig.~\ref{phaseX}~b), for $x=1$, since  for $x=0$ we recover the SSH phase diagram presented in Fig.~\ref{Phase}~a). The main difference between the phase diagrams obtained for $x=0$ in Fig.~\ref{Phase}~a) and for $x=1$ in Fig.~\ref{phaseX}~b) is the existence of the additional topological phase with $\mathcal{W}=-1$ (green) in the latter case. As a consequence, the multi-critical point $(x, t_{1},t_{2})=(1,1,1)$ in Fig.~\ref{phaseX}~b) locates a TPT between three topological regions with $\mathcal{W}=0,\,-1$ and $1$. It is very interesting that by tuning  the parameter $x$ one obtains TPTs in different universality classes, as we discuss in detail below.

\subsection{Penetration depth and Berry connection}

The first point that naturally arises, concerns the validity of the penetration depth approach for a dynamic critical exponent $z \neq1$. In order to elucidate this, we study the decay of the edge states of the SSH model in the presence of a synthetic potential $V(x)$ for $x=0$ and $x=1$.

\begin{figure}[t!]
  \centering
  \includegraphics[width=1\columnwidth]{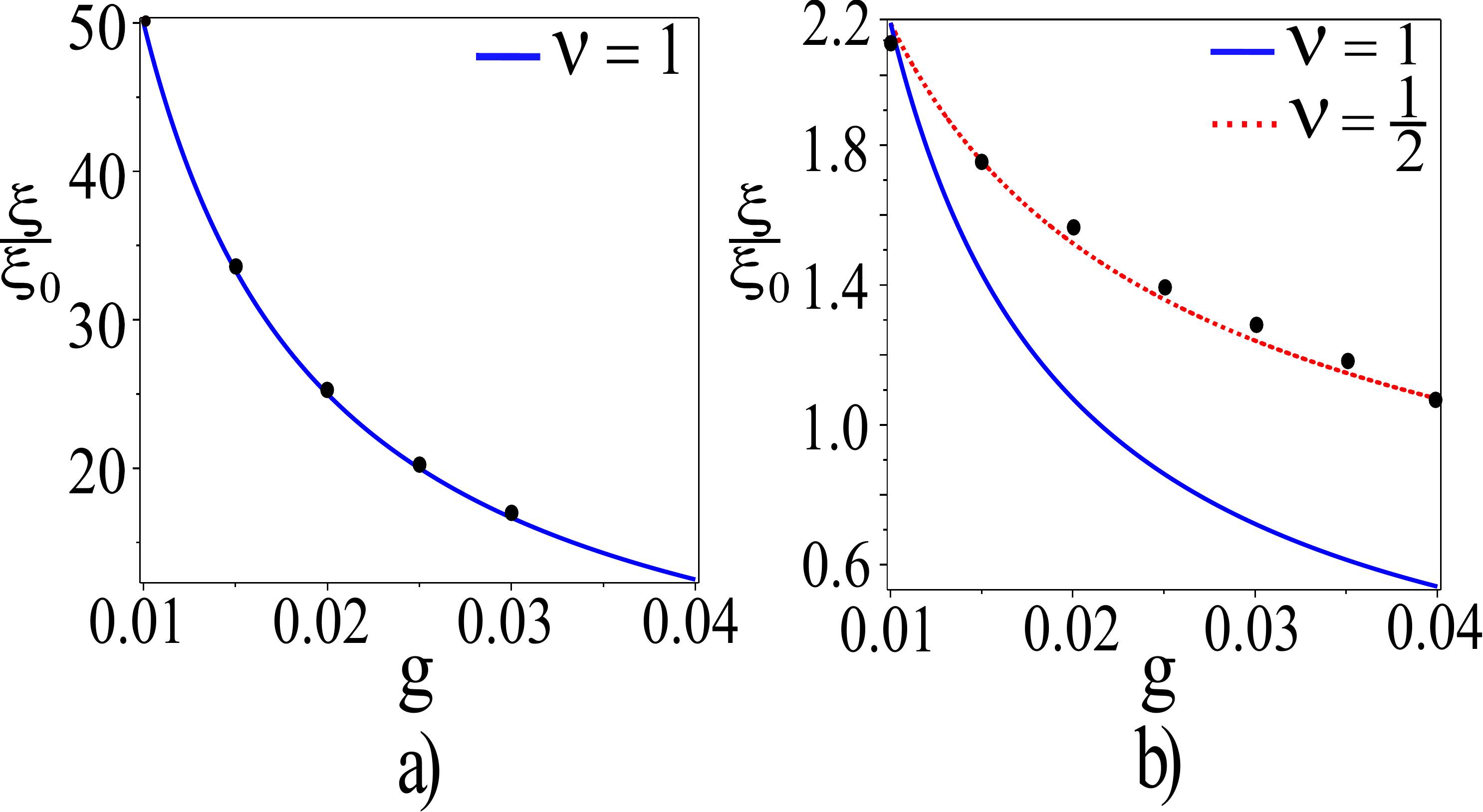}
  \caption{(Color online) Correlation length (penetration depth)  as a function to the distance to the critical point  $g=t_2-t_ {2c}$, with $t_{1}=1$,  for the modified SSH model in the presence of $V(x)$ (black circles). The fittings with the expression $\xi=\xi_0|g|^{-\nu}$ correspond to the solid and dashed lines. The cases $x=0$ and $x=1$ are shown in a) and b), respectively. In a), with $x=0$, the best fit yields $\nu=1$ (blue solid). On the other hand in b), with $x=1$, the best fit is obtained for $\nu=1/2$ (red dashed line). In this case, we also add a tentative fit with  $\nu=1$ (blue solid line) for comparison.}\label{ChenFitPDmeio}
\end{figure}

As in Sec.~\ref{PDepth}, we obtain the edges states of the SSH model in real space for $N=1600$. Notice that now the edge states also depend on the control parameter $x$. The results are presented in Fig.~\ref{ChenFitPDmeio}. In the absence of the synthetic potential, i.e., for $x=0$, the fit of  Fig.~\ref{ChenFitPDmeio}~a) yields  $\nu=1$, as expected. The case for $x=1$ is shown in Fig.~\ref{ChenFitPDmeio}~b), where the fitting of the decay of the edge state in the bulk is well described by a correlation length critical exponent $\nu=1/2$. For comparison we also show in this figure the fitting with $\nu=1$.

As concerns the Berry connection, using an analytical approach, we find that its scaling form up to order $\delta k^4$ is given by,
\begin{equation}\label{Lor4}
F(k,g)= \frac{F(k_0,g)}{1+\left(\frac{A_{2}(x)}{g^{2}}\right)\delta k^2+\left(\frac{A_{4}(x)}{g^{2}}\right)\delta k^4},
\end{equation}
with $\delta k =k-k_{0}$, $k_0 =\pi$, $g=(t_{1}-t_{2})$ and $F(k_{0},g)=(t_{2}-x){g}^{-1}$.  $A_{2}(x)$ and $A_{4}(x)$ are the same coefficients as in  Eq.~(\ref{quadratic}).

\begin{figure}
  \centering
  \includegraphics[width=0.9\columnwidth]{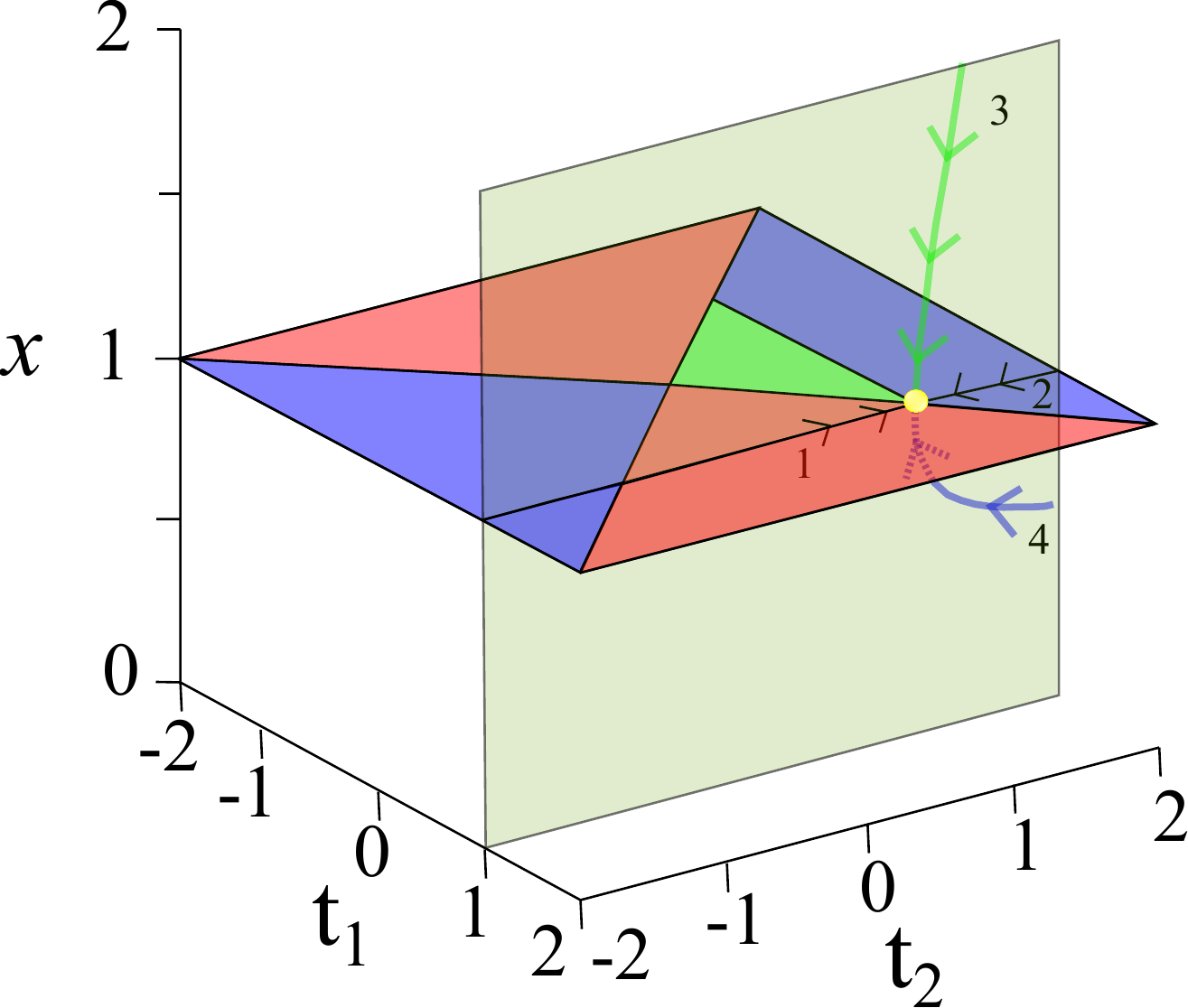}
  \caption{(Color online) The same phase diagram of Fig.~\ref{phaseX} in perspective of the axis $x$. The multi-critical point is located by the (yellow) circle at $(x, t_{1},t_{2})=(1,1,1)$. The trajectories $1$ and $2$, along the $x=1$ plane, correspond to those in Eq.(\ref{Lor4}), characterized respectively by the Berry connection and the penetration depth. The approach to the multi-critical point along the trajectories described by Eq.(\ref{Lor44}) is along the plane $t_{1}=1$ (energy scale). Paths $3$ and $4$ correspond to the two solutions for $A_{2}(x)=0$, namely $x=t_{2}\pm \sqrt{t_{2}^{2}-t_{1}t_{2}}$, respectively. The color scheme of the trajectories $3$ (green) and $4$ (purple) indicate the winding number, as in Fig.~\ref{phaseX}, i.e., $\mathcal{W}=-1$ and $\mathcal{W}=1$, respectively.}\label{Trajetoria}
\end{figure}

Notice that we will fix the energy scale in our problem from now on, taking the hopping $t_1 \equiv 1$. The reason for writing Eq.~(\ref{Lor4}) in this form  is that the quantum multi-critical point $(x, t_1, t_2)=(1,1,1)$ in Fig.~\ref{Trajetoria} can be approached in different ways, as shown in this figure.

\subsection{Approach along the plane $x=1$}

Let us  consider initially fixing $x=1$, such that the approach to the multi-critical point  $(1,1,1)$  is along this plane, as shown by trajectory $1$ in Fig.~\ref{Trajetoria}.
For $x=1$ fixed, we can check that $A_{2} (x=1)/ (g)^{2} = (1 - t_2)^{-1}$,  $A_{4} (x=1)/ (g)^{2} = (1/4)(1 - t_2)^{-2}$ and $F(k_{0}, g)=-1$  in Eq.~(\ref{Lor4}). Furthermore, as $t_2 \rightarrow 1$, $\xi = |g|^{-1/2}$.
Then, when approaching the critical point along the trajectory $x=1$, $t_2 \rightarrow 1^{-}$ from the trivial topological phase with $\mathcal{W}=0$ (red), the Berry connection can be written as,
\begin{equation}\label{Lor4n}
F(k,g)= \frac{-1}{1+ \xi^{2} \delta k^2+ \frac{1}{4}\xi^4\delta k^4},
\end{equation}
where $\xi=1/\sqrt{|1-t_2|}$ that allows to immediately identify the correlation length exponent $\nu=1/2$. Notice that for $k=\pi$, the Berry
connection $F(k=\pi,g)=-1$ is finite.

Next, we approach the critical point still along the plane $x=1$, but with $t_2 \rightarrow 1^{+}$  (trajectory 2 in Fig.~\ref{Trajetoria}). Along this path, we reach the multi-critical point from a non-trivial topological phase with edge modes. This gives the possibility of checking the penetration depth approach as a numerical method for determining the critical exponent $\nu$. The results are shown in Fig.~\ref{ChenFitPDmeio} and confirm the reliability of this procedure for obtaining the correlation length critical exponent, even for a non-Lorentz invariant spectrum.

\subsection{Approach from out of the plane $x=1$}

In this case we take $A_2(x)=0$ in Eqs.~(\ref{quadratic}) and~(\ref{Lor4}). This condition yields two solutions, $x_{\pm}=t_{2}\pm \sqrt{t_{2}^2-t_{1}t_{2}}$.  As $t_{2} \rightarrow 1^{+}$, $x \rightarrow 1$ along the two different trajectories $3$ and $4$, from above ($x_{+}$) or below ($x_{-}$) the plane $x=1$, respectively, as shown in Fig.~\ref{Trajetoria}. Notice that $t_{1} \equiv 1$.
The system along the paths $3$ and $4$ of Fig.~\ref{Trajetoria} are non-trivial topological insulators characterized by the winding numbers $\mathcal{W}=-1$ ($x_{+}$) and $\mathcal{W}=1$ ($x_{-}$), respectively.

Finally, as the multi-critical point is approached along trajectories $3$ and $4$, where $A_2(x_{\pm})=0$, $A_4(x_{\pm})=t_{2}/4$,   the Berry connection is given by,
\begin{equation}\label{Lor44}
F(k,g)=\frac{\mp \sqrt{t_{2}} \xi}{1+\frac{t_2}{4}\xi^{4}\delta k^4},
\end{equation}
\noindent which for $k=\pi$ diverges like the correlation length $\xi=\left|g\right|^{-1/2}=1/\sqrt{|1-t_2|}$. We point out that we also used a numerical approach to obtain the penetration depth along the paths $3$ and $4$ and confirmed the behavior of the correlation length as obtained using the Berry connection.

It is interesting to note that along the trajectories $1$, $3$ and $4$, we observe three different topological phases with $\mathcal{W}=0,-1$ and $1$, respectively, that converge at the multi-critical point ($1,1,1$), the (yellow) circle in Fig.~\ref{Trajetoria}.

\begin{figure}[t!]
  \centering
  \includegraphics[width=1\columnwidth,height=0.6\columnwidth]{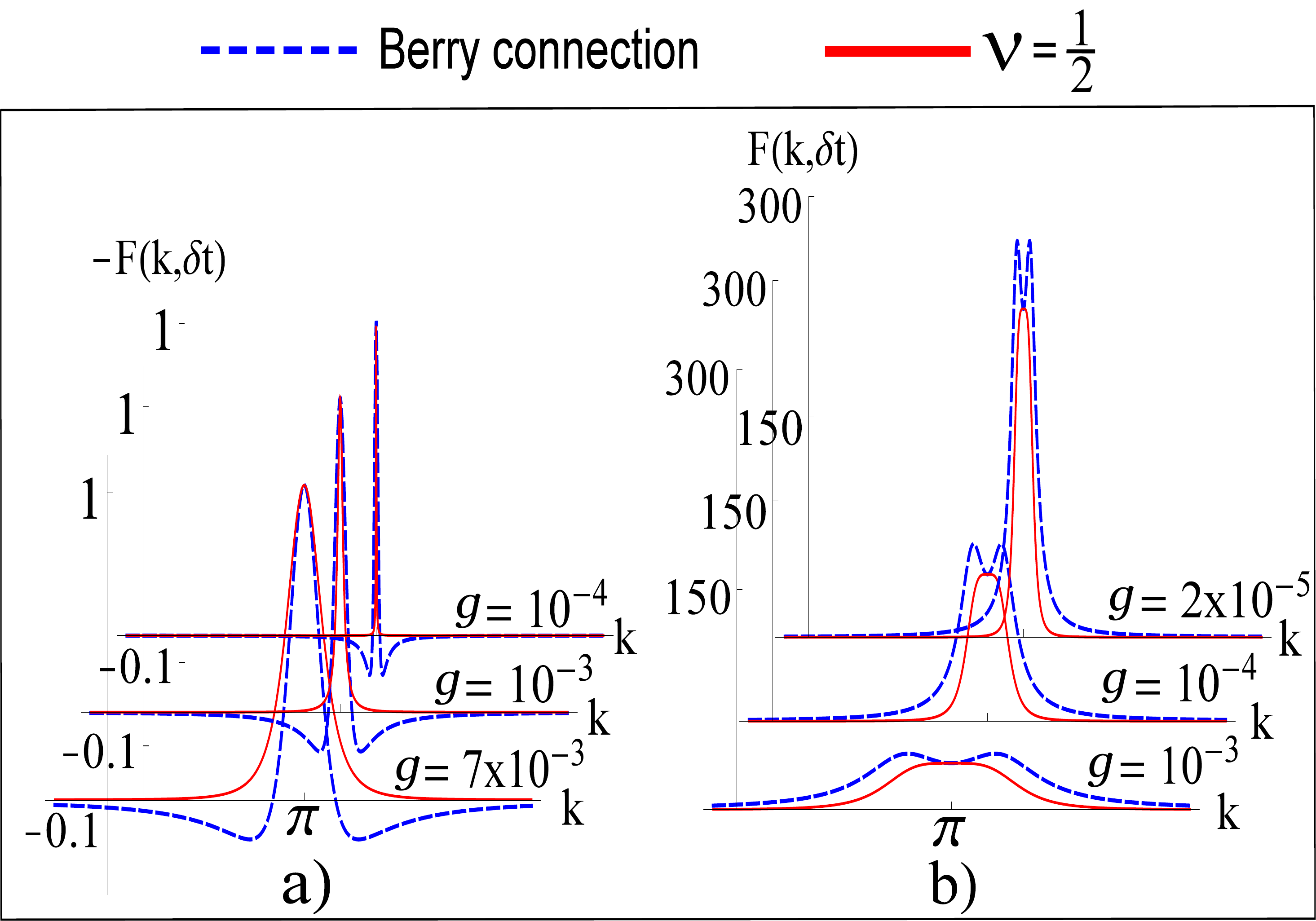}
  \caption{(Color online) Berry connection approach for the modified SSH model in the presence of $V(x)$. The Berry connection and Lorentzian approach for the trajectories described by Eq.(\ref{Lor4}) and Eq.(\ref{Lor44}), are represented in a) and b), respectively. For fixed $x=1$, the Berry connection does not diverge and has a maximum at the gap-closing point $k_{0}=\pi$. However, the width of the Lorentzian, proportional to $1/\xi$, goes to zero as we approach the QCP, signaling the divergence of  $\xi$  as  can be seen in a). On the other hand, in b) for the trajectories along $x=t_{2}\pm \sqrt{t_{2}^{2}-t_{1}t_{2}}$ the divergence of the Berry connection  is restored. In both cases, the trajectories lead to $\nu=1/2$ and $z=2$.}\label{Chenx1xgeral}
\end{figure}

We proceed in Fig.~\ref{Chenx1xgeral} where we present the full Berry connection and the Lorentzian plots for the different trajectories in Fig.~\ref{Trajetoria}. For fixed $x=1$, the trajectory $1$ has the Berry connection investigated  in Fig.~\ref{Chenx1xgeral}~a).

Even though the Berry connection does not diverge along this path ($|F(k_{0},g|\rightarrow1$), we notice that the width of the Lorentzian ($\propto 1/\xi$) close to the critical point goes to zero, signaling the divergence of  $\xi$. So, by fitting an adequate scaling form of the Berry connection as proposed by Chen et al.~\cite{Chen1}, we can determine the correlation length critical exponent.

For trajectories $3$ and $4$, in Fig.~\ref{Chenx1xgeral}~b) the scaling form of the Berry connection, Eq.~(\ref{Lor44}), yields the expected divergence at the QCP as in the Lorentz invariant case of Fig.~\ref{ChenFit}.

We have shown in this Section that even for a non-Lorentz invariant  spectrum, the approach of Chen et al.~\cite{Chen1,Chen2} yield results  for the exponent $\nu$ consistent with the direct numerical determination through the penetration depth of the edge state wave-function.

The multi-critical character of the point $(1,1,1)$ in Fig.~\ref{Trajetoria}  is evidenced by the different behaviors of the Berry connection depending on the path this point is approached.  If we reach this point through trajectory $1$ of Fig.~\ref{Trajetoria}, the Berry connection does not diverge, as shown in Fig.~\ref{Chenx1xgeral} a) or more directly in Eq.~(\ref{Lor4n}). On the other hand, as one approaches this point from out of the plane, as in trajectories 3 or 4 of   Fig.~\ref{Trajetoria}, the Berry connection diverges, as can be seen from Eq.~(\ref{Lor44}) or in Fig.~\ref{Chenx1xgeral} b).
Also consider the line $x=t_1=t_2$ and an arbitrary plane $x=1$ crossing this line at the point $(1,1,1)$, as shown in Fig.~\ref{Trajetoria}. As one approaches this point along this line, Eq.~(\ref{parameter}) is satisfied and we have $z=2$. Notice  that if one approaches this point within the plane  $x=1$ one also gets $z=2$. On the other hand all critical lines in this plane  are associated with a dynamic exponent $z=1$, except the points that satisfy Eq.~(\ref{parameter}). Then, for an arbitrary $x=x_0$, there is always a point in this plane that has $\nu=1/2$ and $z=2$. Note that, the size of the region with $\mathcal{W}=-1$ (green) in Fig.~\ref{phaseX}, or equivalently in Fig.~\ref{Trajetoria}, varies accordingly to the intensity of the synthetic potential. In the renormalization group language, this point, in fact the line ($x=t_1=t_2$) is a line of fully unstable fixed points with critical exponents $\nu=1/2$ and $z=2$. All critical lines that emanate from it have their critical behavior governed by other fixed points with exponents $\nu=1$ and $z=1$.

Chen et al. studied in Ref.~\cite{Chen1} and Ref.~\cite{Chen2}, a specific Hamiltonian in AIII class, given by $H(k)=k^{n}\sigma_{x}+M\sigma_{y}$ (Eq.~131 from Ref.~\cite{Chen2}). They have obtained, using the Berry connection integral that yields the topological invariant, the relation $\nu=1/\Delta \mathcal{W}$. Although our model even in the presence of the synthetic potential still belongs to the same topological class AIII, the model Hamiltonians we have studied here are beyond  Eq.131 of Ref.~\cite{Chen2}.  In our case, the Hamiltonian that leads to Eq.~(\ref{quadratic}), can be written as $H=(a+bk^{2})\sigma_{x}+(ck+dk^{3})\sigma_{y}$, where $a=t_{1}-t_{2}$, $b=t_{2}/2$, $c=-(t_{2}-x)$ and $d=(t_{2}-x)/6$. So, differently from Eq.131 of Ref.~\cite{Chen2}, the coefficients of all Pauli matrices  are now $k$-dependent. As shown  by our study of the phase diagram of Fig.\ref{Phase}, the relation $\nu = 1/\Delta \mathcal{W}$ does not apply in this case.

\section{Conclusions}\label{Conc}

The theory of critical phenomena  is one of the most successful in physics. Recently, a new class of transitions has been discovered that does not conform to the usual paradigms of phase transitions. Topological transitions do not have a clear order parameter and do not present a symmetry breaking.  However,  the existence of a characteristic length that diverges at the topological transition  still allows to use many of the tools of the theory of critical phenomena, as scaling ideas and the renormalization group. The diverging length has been identified as the penetration depth of the surface modes that exist in any non-trivial topological phase. Its divergence is characterized by a critical exponent $\nu$ that obeys  scaling relations and together with the dynamic exponent $z$ and the dimensionality $d$ of the system characterize the universality class of the topological quantum phase transition.

In order to get a deeper insight into this problem, we studied in this paper the critical and topological behavior of  two types of generalized SSH models. The first  with next next nearest neighbors hopping terms and  the second with just nearest neighbors hopping, but in the presence  of a synthetic potential. These models have a rich phase diagram with many topological phases that can be fully characterized by their topological invariants. In spite of this complexity, they  still preserve some simplicity that allows for a thorough analytical examination of their critical and topological properties.

The energy dispersion of the critical modes at a TPT contains information about its critical exponents, as the gap exponent $\nu z$ and the dynamic exponent $z$, as can be seen directly from Eq.~(\ref{Gap}). However, it is crucial to develop general methods to deal with topological phase transitions in any dimension and with anisotropic dispersions. One such approach is a direct numerical calculation of the penetration depth of the surface modes, as we presented here. Another, also investigated in the present work, is based on the scaling properties of the Berry connection~\cite{Chen1,Chen3,Chen4,Chen2}. This relies on the assumption that the Berry connection is a kind of correlation function and as such contains information about the correlation length.

Both techniques allow to obtain the critical exponents. The Berry connection approach may be more suitable if we wish to deal with systems in dimensions larger than one ($d>1$) and with translational symmetry. On the other hand, the penetration depth method may become limited for $d>1$, for computational reasons. Nevertheless, for $d=1$ the penetration depth method allows to perform calculations in systems without translational symmetry.  This means that it can be extended to treat systems with  defects and disordered potentials.

Our extended SSH models constitute an excellent platform to investigate and compare these approaches. We found that both approaches yield the same values for the correlation length exponents in every case studied. In the case of the SSH model with long range interactions, the spectra close to the transitions are Lorentz invariant.  In spite that this model has a rich phase diagram with many phases characterized by different values of the topological invariant, the topological transitions between these phases fall in the same universality class with   correlation length exponent $\nu=1$ and dynamic exponent $z=1$, consistent with the gap exponent $\nu z=1$.

Our results show that the correlation length  exponent of our model, obtained either from a direct numerical calculation of the penetration depth or from a scaling analysis of the Berry connection is independent of the jump of the topological invariant $\mathcal{W}$ at the transition, as discussed in Refs.~\cite{Chen1,Chen2}. The main reason is that the Hamiltonians explored in our paper are beyond that studied by Chen et al. in these references.

The nearest neighbor SSH model in the presence of a synthetic potential $V(x)$ allowed us to study a  topological transition in a model with a spectrum that is not Lorentz invariant. It presents a topological transition in the universality class of the Lifshitz transitions with dynamic critical exponent $z=2$~\cite{Volovik,Volovik2,Imada}. Our results show that for these transitions the correlation length exponent obtained either by a direct numerical calculation of the penetration depth or by the scaling of the Berry connection takes the value $\nu=1/2$, different from the value $\nu=1$ of the Lorentz invariant cases. The model presents a line of multi-critical points at which the value of the Berry connection depends on the path it is approached.

\section{Acknowledgements}

We would like to thank the financial support of the Brazilian Research Agencies CNPq, CAPES and FAPERJ. N.L. would like to thank the CNPq for a doctoral fellowship.

\end{document}